\documentclass[preprint,showpacs,preprintnumbers,showkeys,amsmath,amssymb]{revtex4}

\usepackage{graphicx}
\usepackage{dcolumn}
\usepackage{bm}

\begin{document}


\title{Extinction theorem and propagation of electromagnetic waves between two semi-infinite anisotropic
magnetoelectric materials}

\author{Weixing Shu$^{1}$}\thanks{Corresponding author. $E$-$mail$ $address$: wxshuz@gmail.com.}
\author{Zhongzhou Ren$^{1}$}
\author{Hailu Luo$^{1}$}
\author{Fei Li$^{1}$}
\author{Qin Wu$^{2}$}

\affiliation{$^{1}$Department of Physics, Nanjing University,
Nanjing 210008, China \\ $^{2}$ School of Basic Medical Science,
Guangdong Medical College,  Dongguan 523808, China}

\begin{abstract}
Based on molecular optics we investigate the reflection and
refraction of an electromagnetic wave between two semi-infinite
anisotropic magnetoelectric materials. In terms of Hertz vectors
and the principle of superposition, we generalize the extinction
theorem and derive the propagation characteristics of wave. Using
these results we can easily explain the physical origin of
Brewster effect. Our results extend the extinction theorem to the
propagation of wave between two arbitrary anisotropic materials
and the methods used can be applied to other problems of wave
propagation in materials, such as scattering of light.
\end{abstract}

\maketitle

\section{Introduction}
In classical electromagnetism there are two well-known approaches
to the propagation of electromagnetic waves. The first is to solve
Maxwell's equations with boundary conditions and the second is to
use Ewald-Oseen extinction theorem of molecular optics theory
\cite{Born}. Compared to the former used traditionally, the latter
can give much deeper physical insights into the interaction of
electromagnetic wave with material
\cite{Born,Feynman1963,Wolf1972,Sein1972,Fearn1996}. Using the
extinction theorem, the propagations of electromagnetic waves
through a semi-infinite isotropic material
\cite{Wolf1972,Reali1982} and an isotropic slab \cite{Lai2002}
have been studied. Recently the Brewster mechanism is explained
for light incident onto an isotropic material with negative index
\cite{Fu2005}. The extinction theorem also plays a key role in
light scattering theory \cite{Kong2001}.

In the previous works using the extinction theorem, most deal with
the propagation of electromagnetic waves incident from free space
into isotropic materials. However, the opposite situation from an
material into free space and especially the situation between two
materials is rarely studied. Then, whether the methods used in the
previous work are applicable to the two situations? To answer this
question, it is necessary to generalize the extinction theorem to
investigate the propagation between two materials. On the other
hand, a recent advent of artificial materials, named as
negative-refraction materials, arouses many interest in the field
of optics
\cite{Veselago1968,Shelby2001,Pendry2000,Lindell2001,Smith2004,Luo2002,Hu2002,
Lakhtakia2004,Lakhtakia2006,Lu2004,Luo2005,Luo2006a,Luo2006b,Shen2006}.
Since the negative-refraction materials are actually anisotropic
magnetoelectric, it is also necessary to extend the molecular
optics from isotropic materials to anisotropic materials.

It is the purpose of this letter to use molecular optics to
investigate systematically the propagation of electromagnetic wave
between two semi-infinite anisotropic magnetoelectric materials.
In terms of Hertz vectors and the principle of superposition we
derive the properties of propagation and generalize the extinction
theorem, so that the propagation between two arbitrary materials
can be investigated in a unified framework of molecular optics. We
also explain the mechanism of Brewster effect. The results extend
the conclusions about the propagation in isotropic materials
\cite{Fu2005,Lai2002,Reali1982} and throw light on those obtained
by using Maxwell equations \cite{Shen2006}. The methods used do
not require boundary conditions, but can reveal the interaction of
light with materials, and avoid the difficulties in integrations
and complex calculations usually encountered in using extinction
theorem \cite{Wolf1972}. So the methods here can be applied to
other problems of wave propagation in materials, such as
scattering of light.

\section{Reflection, refraction, and extinction theorem }
In this section, we first employ the formulation of Hertz vector
and the principle of superposition to deduce the radiated fields
generated by dipoles in the propagation of waves between two
anisotropic dielectric-magnetic materials. Then we derive the real
reflected and transmitted fields and the Fresnel's coefficients.
At the same time, we generalize the Ewald-Oseen extinction
theorem. Throughout the paper SI units are used.

\begin{figure}[t]
\centering
\includegraphics[width=8cm]{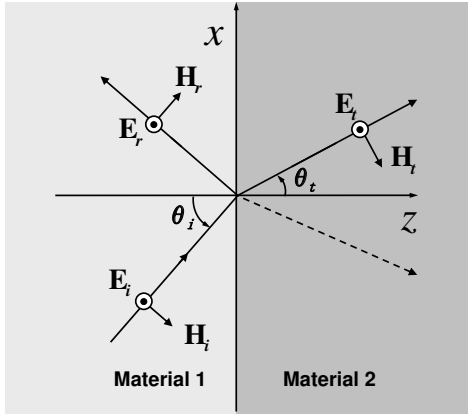}
\caption{\label{shuFig1}Schematic diagram for the reflection and
refraction of TE waves at the interface between vacuum and an
anisotropic material. The dashed line denotes the possible ray of
negative refraction.}
\end{figure}

Let us consider a monochromatic electromagnetic field of ${\bf
E}_i={\bf E}_{i0} \exp{(i{\bf k}_i\cdot {\bf r}-i\omega t)}$ and
${\bf H}_i={\bf H}_{i0} \exp{(i{\bf k}_i\cdot {\bf r}-i\omega t)}$
incident from an anisotropic material into another one filling the
semi-infinite space $z>0$ with ${\bf k}_i=k_{ix}\hat{{\bf
x}}-k_{iz}\hat{{\bf z}}$. The schematic diagram is in
Fig.~\ref{shuFig1}. Since the material responds linearly, all the
fields have the same dependence of $\exp{(-i\omega t)}$ which will
be omitted subsequently. For simplicity, the permittivity and
permeability tensors of materials are assumed diagonal
simultaneously in the principal coordinate system,
$\boldsymbol{\varepsilon}_j= \hbox{diag}[\varepsilon_{jx},
\varepsilon_{jy}, \varepsilon_{jz}]$,
$\boldsymbol{\mu}_j=\hbox{diag}[\mu_{jx}, \mu_{jy}, \mu_{jz}]$,
$j=1,2$.

Following the molecular optics theory, a bulk material can be
regarded as a collection of molecules (or atoms) embedded in the
vacuum. Driven by the external field, the molecules are brought
into oscillations and then secondary waves are generated by the
induced dipoles. The radiated electric field by dipoles is decided
by \cite{Born}
\begin{eqnarray}\label{E_dipole_field}
&&{\bf
E}_{rad}=\nabla(\nabla\cdot{\bf{\Pi}}_e)-\varepsilon_0\mu_0\frac{\partial^2{\bf{\Pi}}_e}{\partial
t^2}-\mu_0\nabla\times\frac{\partial{\bf{\Pi}}_m}{\partial t}
\end{eqnarray}
and the  generated magnetic field is
\begin{eqnarray}\label{H_dipole_field}
&&{\bf
H}_{rad}=\nabla(\nabla\cdot{\bf{\Pi}}_m)-\varepsilon_0\mu_0\frac{\partial^2{\bf{\Pi}}_m}{\partial
t^2}+\varepsilon_0\nabla\times\frac{\partial{\bf{\Pi}}_e}{\partial
t}.
\end{eqnarray}
Here ${\bf \Pi}_e$ and ${\bf \Pi}_m$ are the Hertz vectors,
\begin{eqnarray}\label{Pi_e}
&&{\bf \Pi}_e({\bf r})=\int \frac{{\bf P}({\bf
r'})}{\varepsilon_0}G({\bf r}-{\bf r}')\hbox{d}{\bf r}',\\
&&{\bf \Pi}_m({\bf r})=\int {\bf M}({\bf r'})G({\bf r}-{\bf
r}')\hbox{d}{\bf r}'.\label{Pi_m}
\end{eqnarray}
The dipole moment density of electric dipoles ${\bf P}$  and that
of magnetic dipoles ${\bf M}$ are related to the associated field
as ${\bf P}=\varepsilon_0\boldsymbol{\chi}_e\cdot{\bf E}$, ${\bf
M}=\boldsymbol{\chi}_m\cdot{\bf H}$, where the electric
susceptibility
$\boldsymbol{\chi}_e=(\boldsymbol{\varepsilon}/{\varepsilon_0})-1$
and the magnetic susceptibility
$\boldsymbol{\chi}_m=(\boldsymbol{\mu}/{\mu_0})-1$. The Green
function is $G({\bf r}-{\bf r}')=\exp{(ik_0|{\bf r}-{\bf
r}'|)}/(4\pi|{\bf r}-{\bf r}'|)$, where $k_0$ is the wave number
in vacuum. In the first medium the dipoles produce forward waves
as well as backward waves. So, we assume the associated dipole
moment densities have the following forms
\begin{eqnarray}\label{PM1} {\bf P}_{1}&=&{\bf
P}_{1F}\exp{(i{\bf k}_{1F}\cdot {\bf r})}+{\bf
P}_{1B}\exp{(-i{\bf k}_{1B}\cdot {\bf r})},\nonumber\\
{\bf M}_{1}&=&{\bf M}_{1F}\exp{(i{\bf k}_{1F}\cdot {\bf r})}+{\bf
M}_{1B}\exp{(-i{\bf k}_{1B}\cdot {\bf r})},
\end{eqnarray}
while ${\bf P}_{2}={\bf P}_{2F}\exp{(i{\bf k}_{2F}\cdot {\bf r})}$
and ${\bf M}_{2}={\bf M}_{2F}\exp{(i{\bf k}_{2F}\cdot {\bf r})}$
in the second material, where $F$ and $B$ label the forward and
backward propagating waves, respectively. We now determine the
Hertz vectors. We firstly represent the Green function in the
Fourier form \cite{Reali1982}. Then, inserting it into
Eq.~(\ref{Pi_e}) and using the delta function definition and
contour integration method, the Hertz vectors can be evaluated
\begin{widetext}
\begin{eqnarray}\label{Pi_1e}
{\bf \Pi}_{1e}=\left\{
\begin{array}{cc}
\frac{{\bf P}_{1F}}{\varepsilon_0}\left [\frac{\exp{(i{\bf
k}_{1F}\cdot {\bf r})}}{k_{1F}^2-q_{1F}^2}+\frac{\exp{(i{\bf
q}_{1B}\cdot {\bf r})}}{2q_{1z}(q_{1z}+k_{1Fz})}\right
]+\frac{{\bf P}_{1B}}{\varepsilon_0}\left[\frac{\exp{(-i{\bf
k}_{1B}\cdot {\bf r})}}{k_{1B}^2-b_{1B}^2}+\frac{\exp{(-i{\bf
b}_{1F}\cdot {\bf r})}}{2b_{1z}(b_{1z}-k_{1Bz})}\right],&z<0\\
-\frac{{\bf P}_{1F}}{\varepsilon_0}\frac{\exp{(i{\bf q}_{1F}\cdot
{\bf r})}} {2q_{1z}(q_{1z}-k_{1Fz})}-\frac{{\bf
P}_{1B}}{\varepsilon_0}\frac{\exp{(-i{\bf b}_{1B}\cdot {\bf r})}}
{2b_{1z}(b_{1z}+k_{1Fz})},&z\geq0
\end{array}\right.
\end{eqnarray}
\end{widetext}
\begin{eqnarray}\label{Pi_2e}
{\bf \Pi}_{2e}=\left\{
\begin{array}{cc}
-\frac{{\bf P}_{2F}}{\varepsilon_0} \frac{\exp{(i{\bf q}_{2B}\cdot
{\bf
r})}}{2q_{2z}(q_{2z}+k_{2Fz})},&z<0\\
\frac{{\bf P}_{2F}}{\varepsilon_0}\left[\frac{\exp{(i{\bf
k}_{2F}\cdot {\bf r})}}{k_{2F}^2-q_{2F}^2}+\frac{\exp{(i{\bf
q}_{2F}\cdot {\bf r})}} {2q_{2z}(q_{2z}-k_{2Fz})}\right],&z\geq0
\end{array}\right.
\end{eqnarray}
where
\begin{eqnarray}
&&{\bf q}_{jF}=k_{jFx}\hat{{\bf x}}+q_{jz}\hat{{\bf z}}, {\bf
q}_{jB}=k_{jFx}\hat{{\bf x}}-q_{jz}\hat{{\bf z}},
q_{jz}^2=k_0^2-k_{jFx}^2,\nonumber\\
&&{\bf b}_{1F}=k_{1Bx}\hat{{\bf x}}+b_{1z}\hat{{\bf z}}, {\bf
b}_{1B}=k_{1Bx}\hat{{\bf x}}-b_{1z}\hat{{\bf z}},
b_{1z}^2=k_0^2-k_{1Bx}^2, j=1,2.
\end{eqnarray}

Following the superposition theory of field, the fields in the
right region produced by the dipoles in the two media add up to
the transmitted field. Then, we have
\begin{equation}\label{E_t=}
{\bf E}_t={\bf E}^{1.right}_{rad}+{\bf E}^{2}_{rad},
\end{equation}
where the contribution from the first medium to the right side,
${\bf E}^{1.right}_{rad}$,  can be calculated by substituting
Eq.~(\ref{Pi_1e}) into Eq.~(\ref{E_dipole_field})
\begin{eqnarray}\label{E^{1.right}_{rad}}
{\bf E}^{1.right}_{rad}=-\frac{{\bf Q}({\bf q}_{1F},{\bm P}_{1F})}
{2q_{1z}(q_{1z}-k_{1Fz})}-\frac{{\bf Q}(-{\bf b}_{1B},{\bm
P}_{1B})} {2b_{1z}(b_{1z}+k_{1Fz})},
\end{eqnarray}
the field radiated by the second medium itself, ${\bf
E}^{2}_{rad}$, can be obtained after inserting Eq.~(\ref{Pi_2e})
into Eq.~(\ref{E_dipole_field})
\begin{eqnarray}\label{E^{2}_{rad}}
{\bf E}^{2}_{rad}&=&\frac{{\bf Q}({\bf k}_{2F},{\bm
P}_{2F})}{k_{2F}^2-q_{2F}^2}+\frac{{\bf Q}({\bf q}_{2F},{\bm
P}_{2F})} {2q_{2z}(q_{2z}-k_{2Fz})},
\end{eqnarray}
and ${\bf Q}$ is an auxiliary function
\begin{equation}
{\bf Q}({\bf K},{\bm P})\equiv-\frac{1}{\varepsilon_0}\left[{\bf
K}\times{\bf K}\times {\bm P}+({\bf
K}^2-\varepsilon_0\mu_0\omega^2){\bm
P}+\omega\mu_0\varepsilon_0{\bf K}\times {\bm M}\right]\exp{(i{\bf
K}\cdot {\bf r})}
\end{equation}
where ${\bf M}=\boldsymbol{\chi}_m\cdot \{{\bf k}\times [{\bf
P}/(\varepsilon_0\boldsymbol{\chi}_e)]/\boldsymbol{\mu}\}/\omega$.
Substituting Eqs.~(\ref{E^{1.right}_{rad}}) and
(\ref{E^{2}_{rad}}) into (\ref{E_t=}), and considering the
transmitted field with the  form
\begin{equation}\label{E_t}
{\bf E}_t= \frac{{\bf
P}_{2F}}{\varepsilon_0\boldsymbol{\chi}_{2e}}\exp{(i{\bf
k}_{2F}\cdot {\bf r})},
\end{equation}
we come to the following conclusions by a self-consistent
analysis. (i). From terms with the phase factor $\exp{(i{\bf
k}_{2F}\cdot{\bf r})}$ in Eq.~(\ref{E_t=}) yield ${\bf
k}_{2F}={\bf k}_{t}$ and the dispersion relation
\begin{equation}
\frac{k_{tx}^2}{\mu_{2z}
\varepsilon_{2y}}+\frac{k_{tz}^2}{\mu_{2x} \varepsilon_{2y}
}=\omega^2, \frac{k_{tx}^2}{\varepsilon_{2z}
\mu_{2y}}+\frac{k_{tz}^2}{\varepsilon_{2x} \mu_{2y} }=\omega^2
\label{D1}
\end{equation}
for TE and TM waves, respectively. (ii). We know that ${\bf
q}_{1F}$, ${\bf q}_{2F}$ and ${\bf b}_{1B}$ are all vacuum wave
vectors. Since only ${\bf k}_{2F}$ appears in the final
transmitted field, they all should be extinguished. So we conclude
that ${q}_{1z}={q}_{2z}=b_{1z}$ and $k_{1Fx}=k_{2Fx}=-k_{1Bx}$.
Then, comes naturally the Snell's law:
$k_{1F}\sin\theta_i=k_{2F}\sin\theta_t$. At the same time,
\begin{equation}\label{extinction_right}
\frac{{\bf Q}({\bf q}_{1F},{\bm
P}_{2F})}{q_{1z}-k_{2Fz}}-\frac{{\bf Q}({\bf q}_{1F},{\bm
P}_{1F})} {q_{1z}-k_{1Fz}}-\frac{{\bf Q}({\bf q}_{1F},{\bm
P}_{1B})} {q_{1z}+k_{1Fz}}=0.
\end{equation}
Equation~(\ref{extinction_right}) is the generalized expression of
the extinction theorem about the forward vacuum waves. It
describes how the radiation field produced by the dipoles of the
first medium is extinguished by the counterpart in the second
medium.

Now we study the fields in the first material. Similarly, all the
fields radiated by the whole space are superposed to form the
incident and reflected field.
\begin{equation}\label{E_i+r=}
{\bf E}_i+{\bf E}_r={\bf E}^{1}_{rad}+{\bf E}^{2.left}_{rad},
\end{equation}
where the field radiated by the first medium, ${\bf E}^{1}_{rad}$,
can be calculated substituting Eq.~(\ref{Pi_1e}) into
Eq.~(\ref{E_dipole_field})
\begin{eqnarray}\label{E^{1}_{rad}}
{\bf E}^{1}_{rad}=\frac{{\bf Q}({\bf k}_{1F},{\bm
P}_{1F})}{k_{1F}^2-q_{1F}^2}+\frac{{\bf Q}({\bf q}_{1B},{\bm
P}_{1F})}{2q_{1z}(q_{1z}+k_{1Fz})}+\frac{{\bf Q}(-{\bf
k}_{1B},{\bm P}_{1B})}{k_{1B}^2-b_{1B}^2}+\frac{{\bf Q}({\bf
q}_{1B},{\bm P}_{1B})}{2q_{1z}(q_{1z}-k_{1Bz})},
\end{eqnarray}
the contribution ${\bf E}^{2.left}_{rad}$ from the second medium
to the left half-space can be obtained after inserting
Eq.~(\ref{Pi_2e}) into Eq.~(\ref{E_dipole_field})
\begin{eqnarray}\label{E^{2.left}_{rad}}
{\bf E}^{2.left}_{rad}&=&-\frac{{\bf Q}({\bf q}_{2B},{\bm
P}_{2F})}{2q_{2z}(q_{2z}+k_{2Fz})},
\end{eqnarray}
The incident and reflected fields may be written as
\begin{equation}\label{E_i=P}
{\bf E}_i= \frac{{\bf
P}_{1F}}{\varepsilon_0\boldsymbol{\chi}_{1e}}\exp{(i{\bf
k}_{1F}\cdot {\bf r})}, {\bf E}_r= \frac{{\bf
P}_{1B}}{\varepsilon_0\boldsymbol{\chi}_{1e}}\exp{(-i{\bf
k}_{1B}\cdot {\bf r})},
\end{equation}
respectively. Inserting Eqs.~(\ref{E^{1}_{rad}}),
(\ref{E^{2.left}_{rad}}) and (\ref{E_i=P}) into (\ref{E_i+r=})
which must hold true for everywhere in the left half-space, we
come to the following conclusions. (i). From terms with the phase
factor $\exp{(i{\bf k}_{1F}\cdot{\bf r})}$ or $\exp{(i{\bf
k}_{1B}\cdot{\bf r})}$ in Eq.~(\ref{E_i+r=}) follow that ${\bf
k}_{1F}={\bf k}_{i}$,  ${\bf k}_{1B}={\bf k}_{r}$ and the
dispersion relation is like Eq.~(\ref{D1}) after replacing the
subscripts 2 with 1 and $t$ with $i$, respectively. (ii). ${\bf
q}_{1B}$ and ${\bf q}_{2B}$  are vacuum wave vectors and should be
extinguished. So, we have ${\bf q}_{1B}={\bf q}_{2B}$ and
\begin{equation}\label{extinction_left}
\frac{{\bf Q}({\bf q}_{1B},{\bm
P}_{1F})}{q_{1z}+k_{1Fz}}+\frac{{\bf Q}({\bf q}_{1B},{\bm
P}_{1B})}{q_{1z}-k_{1Fz}}-\frac{{\bf Q}({\bf q}_{1B},{\bm
P}_{2F})}{q_{1z}+k_{2Fz}}=0,
\end{equation}
which is a new expression of the extinction theorem we find. It
shows how the the backward vacuum field produced by the dipoles of
the second medium is extinguished by that in the first medium.

Solving the set of Eqs.~(\ref{E_t}), (\ref{extinction_right}),
(\ref{E_i=P}), and (\ref{extinction_left}), we obtain the
reflection coefficient $R_E={ E}_{r0}/{ E}_{i0}$ and the
transmission coefficient $T_E={E}_{t0}/{ E}_{i0}$ for TE waves
\begin{equation}\label{RE}
R_E=\frac{\mu_{2x}k_{iz}-\mu_{1x}k_{tz}}{\mu_{2x}k_{iz}+\mu_{1x}k_{tz}}
,~~~T_E=\frac{2\mu_{2x}k_{iz}}{\mu_{2x}k_{iz}+\mu_{1x}k_{tz}}.
\end{equation}
Analogously, we can discuss TM waves and obtain similar results
after replacing $\boldsymbol{\mu}$ in with
$\boldsymbol{\varepsilon}$, considering the expressions of ${\bf
E}$ in Eq.~(\ref{E_dipole_field}) and ${\bf H}$ in
Eq.~(\ref{H_dipole_field}).

Hence, we have obtained the generalized extinction theorem, i.e.
Eqs.~(\ref{extinction_right}) and (\ref{extinction_left}). So, the
propagation between two arbitrary materials can be studied in a
unified framework: Under the action of external field in the first
material, the molecules in the tow materials are driven to
oscillate and generate induced fields. The transmitted wave is the
result of superposition of the incident vacuum field from the
first medium and the fields radiated by the induced dipoles in the
second medium. The reflected wave is the sum of the backward
vacuum field from the second medium and the backward fields
induced by the dipoles in the first medium. Note that
Eqs.~(\ref{extinction_right}) and (\ref{extinction_left}) should
hold true for every point in the associated half-spaces, which
indicates that the cancellation of vacuum waves occurs everywhere
inside the materials.

\section{Origin of Brewster effect}

\begin{figure}
\centering
\includegraphics[width=8cm]{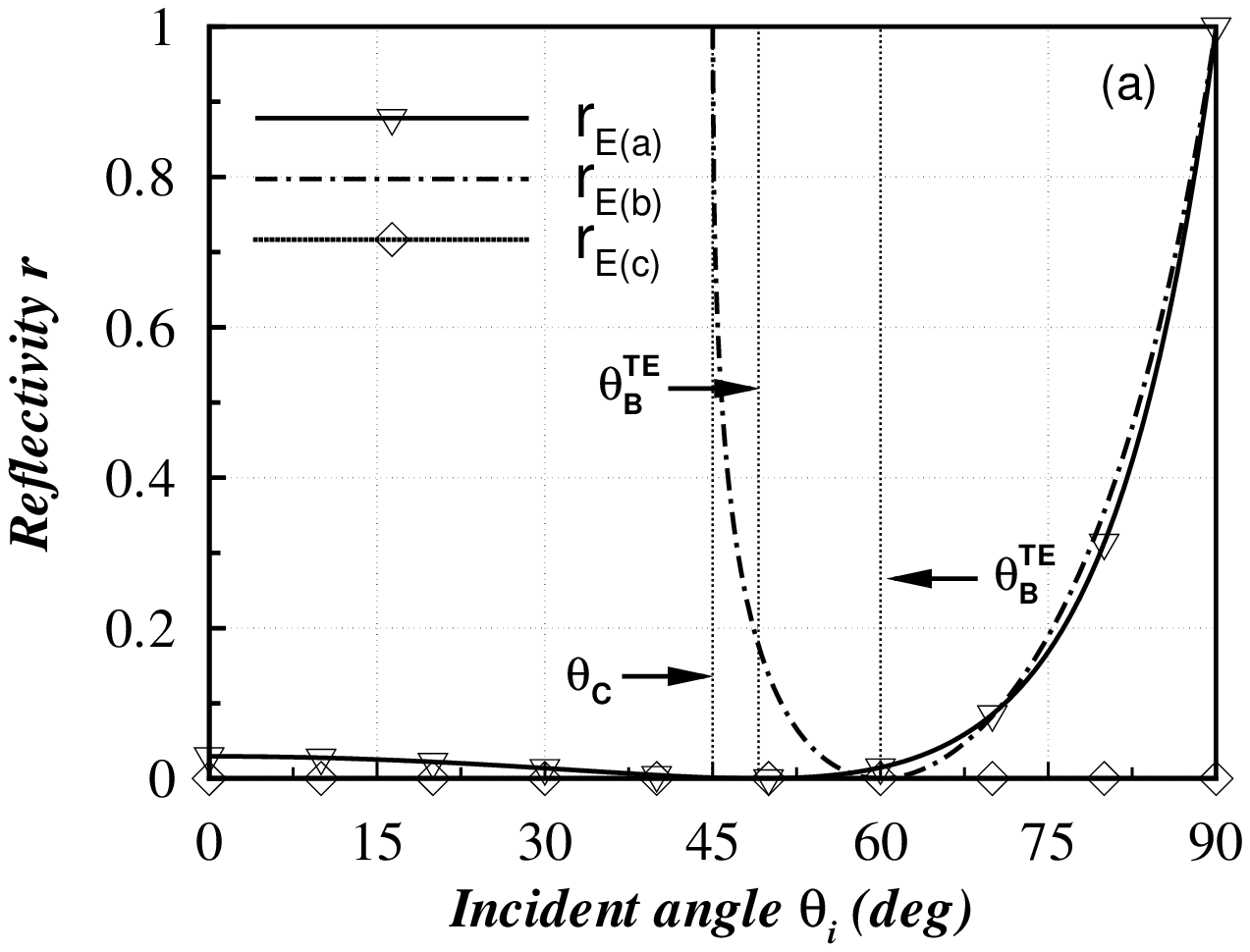}
\includegraphics[width=8cm]{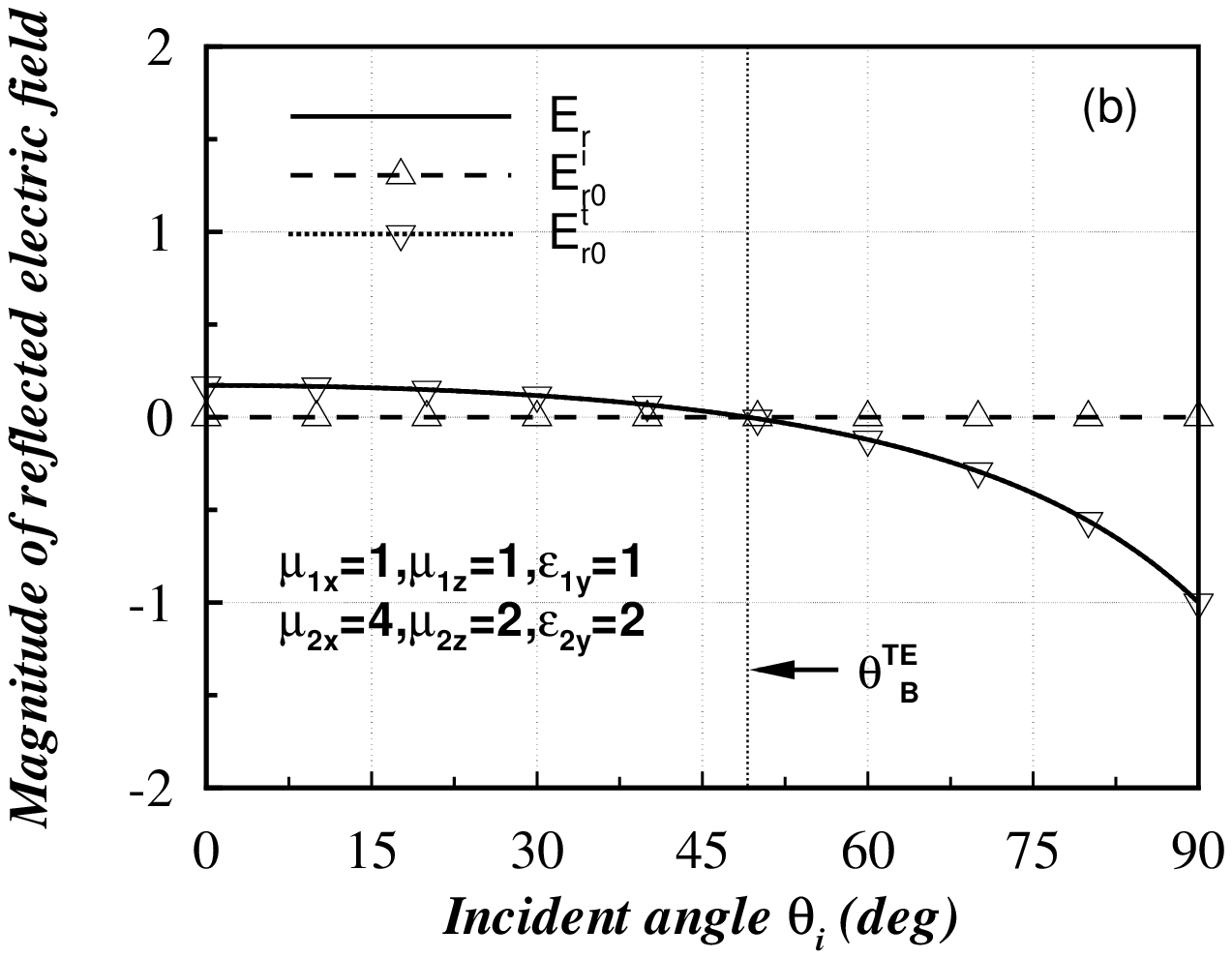}
\includegraphics[width=8cm]{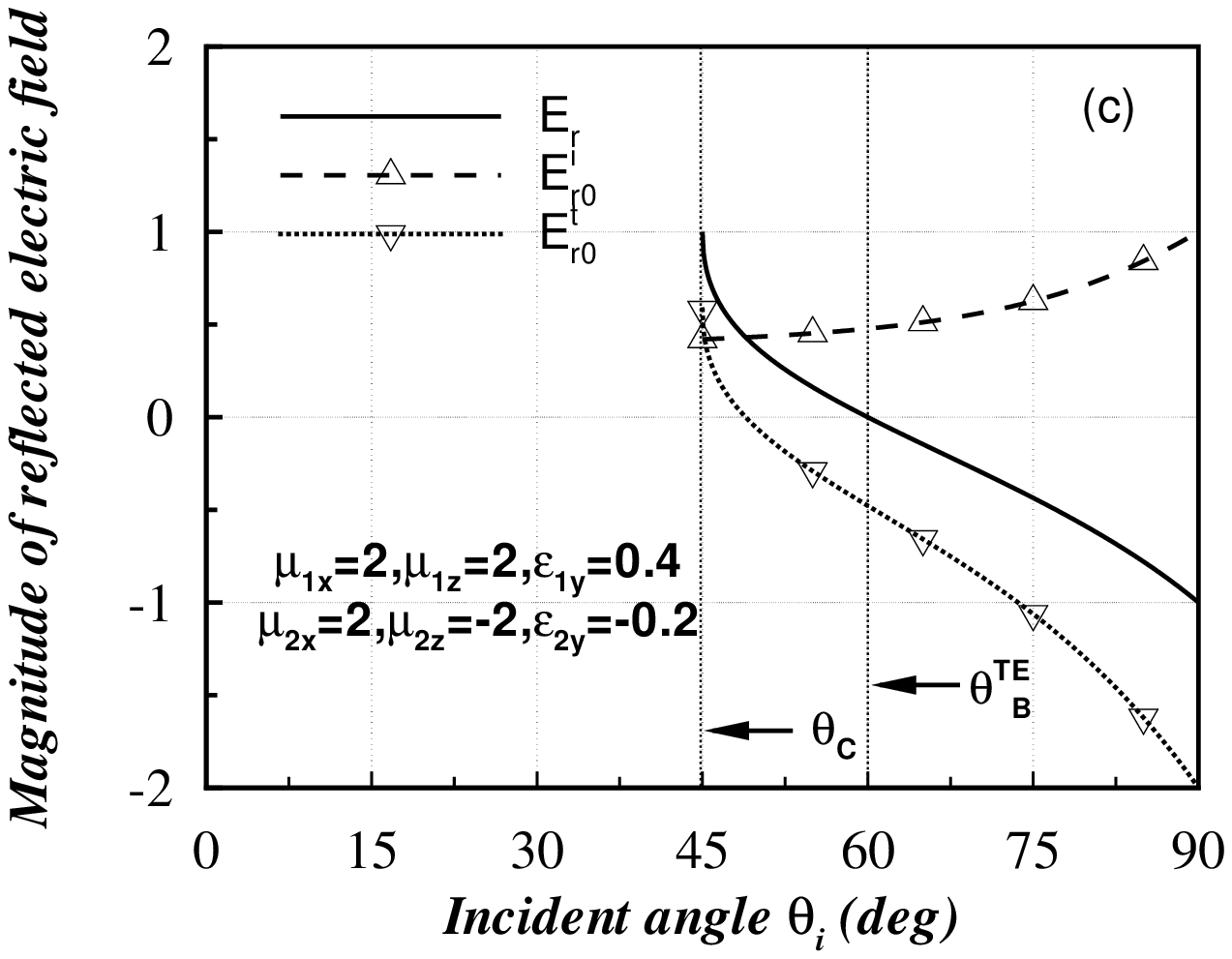}
\includegraphics[width=8cm]{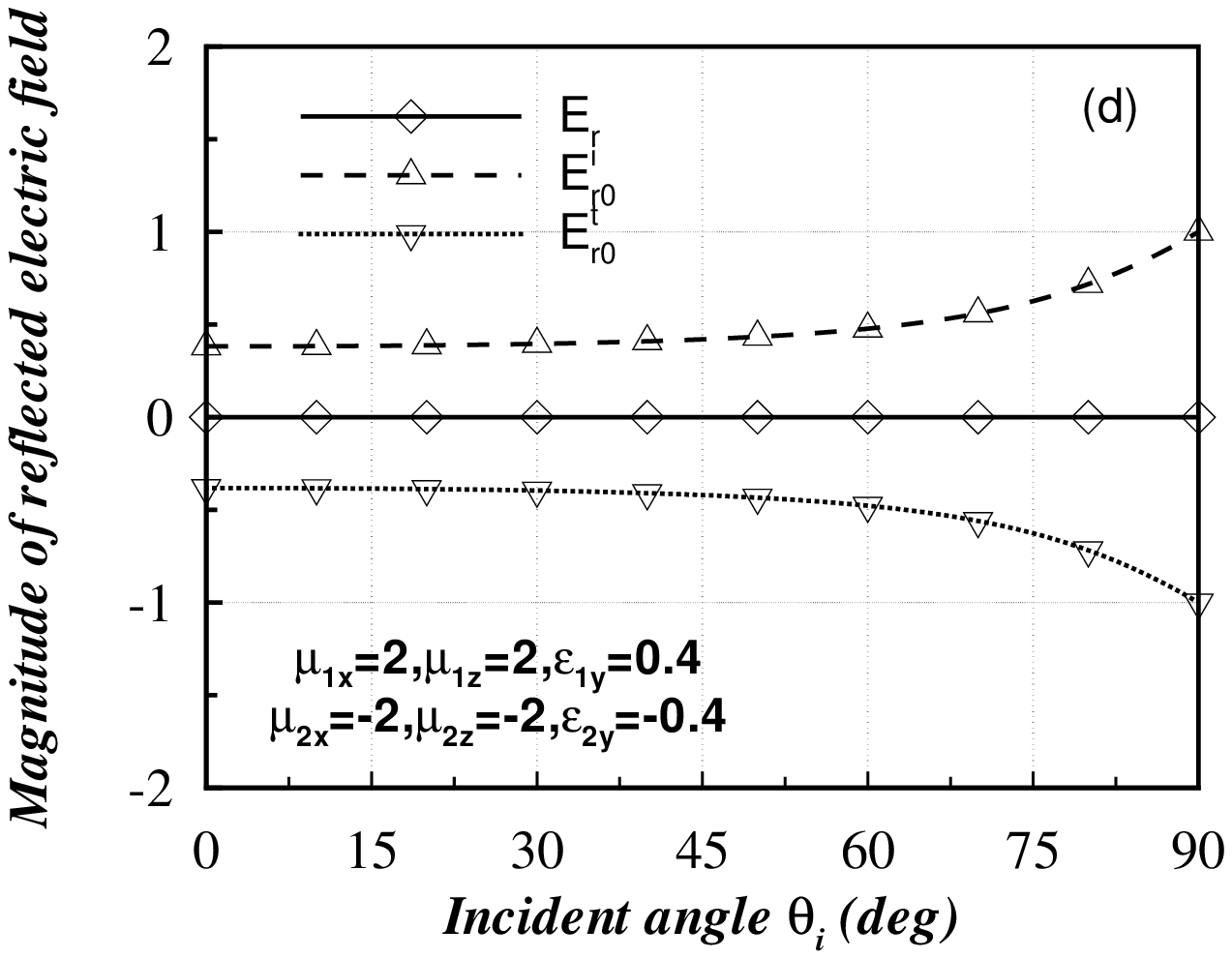}
\caption{\label{cutoff} Reflectivity and reflected field
magnitudes, normalized by the corresponding incident field
magnitudes, for TE wave incident incident on the interface between
two anisotropic materials. The three curves of reflectivity in (a)
correspond the case in (b), (c), and (d), respectively. Since the
first medium is vacuum in (b), the radiated electric fields (${\bf
E}^t_{r0}$) generated by the second medium forms the reflected
field, then Brewster angle $\theta_B^{TE}$ appears when ${\bf
E}^t_{r0}=0$. In (c), $\theta_c$ is the critical angle of
incidence. In (d), ${\bf E}^i_{r0}+{\bf E}^t_{r0}\equiv 0$, so
omnidirectional total transmission occurs.}\label{figfield}
\end{figure}

\begin{figure}
\centering
\includegraphics[width=16cm]{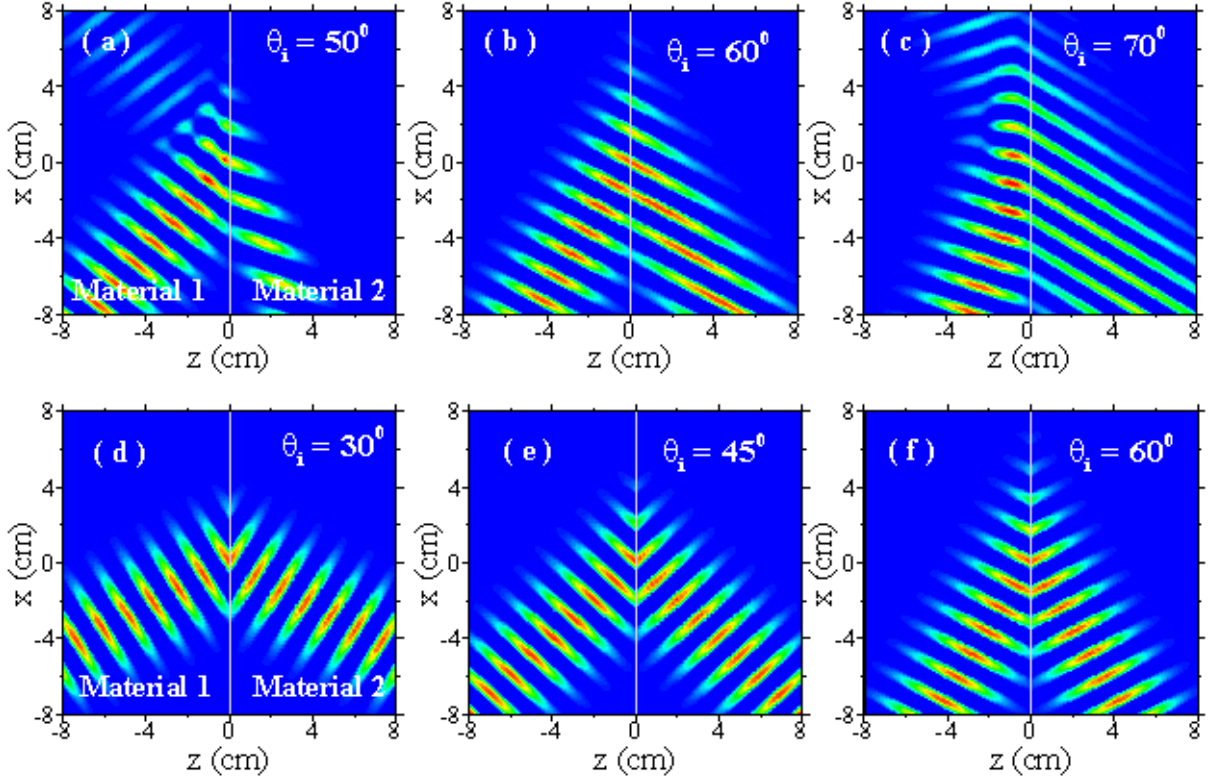}
\caption{\label{cutoff}Reflection and refraction of a Gaussian
beam incident on the interface between two anisotropic materials.
For (a)-(c) and (d)-(f) the two materials correspond to those in
(c) and (d) of Fig.~\ref{figfield}, respectively. The phase and
the energy flow are refracted regularly and anomalously,
respectively in (a)-(c), but are both refracted anomalously in
(d)-(f).}\label{figbeam}
\end{figure}

In what follows we apply the above conclusions to discuss the
origin of Brewster effect. If the power reflectivity $|R|^2=0$,
there is no reflected wave and the incident angle is called
Brewster angle
\cite{Zhou2003,Grzegorczyk2005,Tamayama2006,Kong2000}. From
Eqs.~(\ref{extinction_left}) follows the reflected field magnitude
\begin{eqnarray}\label{E_r=}
{\bf E}_{r0}&=& \frac{1}{q_{1z}+k_{iz}\mu_0/\mu_{1x}}\left
\{\frac{{\bf q}_{1B}\times[{\bf q}_{1B}\times
(\boldsymbol{\chi}_{1e}\cdot{\bf
E}_{i0})]}{q_{1z}+k_{iz}}+\frac{{\bf
q}_{1B}\times\{\boldsymbol{\chi}_{1m}\cdot[{\mu_0}{\boldsymbol{\mu}}_1^{-1}\cdot({\bf
k}_i\times {\bf E}_{i0})]\}}{q_{1z}+k_{iz}}\right.\nonumber\\
&&\left.+\frac{{\bf q}_{1B}\times[{\bf q}_{1B}\times
(\boldsymbol{\chi}_{2e}\cdot{\bf
E}_{t0})]}{q_{1z}+k_{tz}}+\frac{{\bf
q}_{1B}\times\{\boldsymbol{\chi}_{2m}\cdot[{\mu_0}{\boldsymbol{\mu}}_2^{-1}\cdot({\bf
k}_t\times {\bf E}_{t0})]\}}{q_{1z}+k_{tz}}\right\}.
\end{eqnarray}
On the right-hand side of Eq.~(\ref{E_r=}), the first two denote
the contributions (labelled as ${\bf E}^i_{r0}$) of dipoles in the
first medium to the reflected field, while the last two are the
contributions (labelled as ${\bf E}^t_{r0}$) from the second
medium. In order for zero reflection, it requires that ${\bf
E}^i_{r0}+{\bf E}^t_{r0}=0$, from which follows the condition for
Brewster effect: If
\begin{equation}\label{Brewster_E_condition}
\frac{\mu_{2z}\varepsilon_{2y}-\mu_{1z}\varepsilon_{1y}}
{\mu_{1z}\mu_{2z}(\mu_{2x}\varepsilon_{1y}-\mu_{1x}\varepsilon_{2y})}
>0,
\end{equation}
the Brewster angle for TE waves is
\begin{equation}\label{Brewster_E}
\theta_{B}^{TE}=\cot ^{-1}
\sqrt{\frac{\mu_{1x}^2(\mu_{2z}\varepsilon_{2y}-\mu_{1z}\varepsilon_{1y})}
{\mu_{1z}\mu_{2z}(\mu_{2x}\varepsilon_{1y}-\mu_{1x}\varepsilon_{2y})}}\,;
\end{equation}
If
$\mu_{2x}/\mu_{1x}=\varepsilon_{2y}/\varepsilon_{1y}\cap\varepsilon_{2y}/\varepsilon_{1y}\neq\mu_{1z}/\mu_{2z}$,
then $\theta_{B}^{TE}=0$; When
$\mu_{2x}/\mu_{1x}\neq\varepsilon_{2y}/\varepsilon_{1y}\cap\varepsilon_{2y}/\varepsilon_{1y}=\mu_{1z}/\mu_{2z}$,
$\theta_{B}^{TE}=\pi/2$; If
$\mu_{2x}/\mu_{1x}=\varepsilon_{2y}/\varepsilon_{1y}\cap\varepsilon_{2y}/\varepsilon_{1y}=\mu_{1z}/\mu_{2z}$,
then Brewster effect will occur for any angle of incidence, which
may lead to important applications in optics. All the typical and
nontrivial sign combinations of $\boldsymbol{\varepsilon}$ and
$\boldsymbol{\mu}$ are shown in Table.~\ref{Brewster label}.
\begin{table}

\caption{Existence conditions of Brewster angles for TE waves at
the interface of two anisotropic media.~\label{Brewster label}}

\begin{ruledtabular}
\begin{tabular}{ccccccc}
$\mu_{1x}$   &   $\mu_{1z}$ &$\varepsilon_{1y}$ & $\mu_{2x}$
 &    $\mu_{2z}$ & $\varepsilon_{2y}$  & Existence conditions \\ \hline
$+$ & $+$ & $+$ & $+$ & $+$ & $+$ &
$\frac{\mu_{2x}}{\mu_{1x}}>\frac{\varepsilon_{2y}}{\varepsilon_{1y}}\cap
 \frac{\varepsilon_{2y}}{\varepsilon_{1y}}>\frac{\mu_{1z}}{\mu_{2z}}$,
 $\frac{\mu_{2x}}{\mu_{1x}}<\frac{\varepsilon_{2y}}{\varepsilon_{1y}}\cap
 \frac{\varepsilon_{2y}}{\varepsilon_{1y}}<\frac{\mu_{1z}}{\mu_{2z}}$  \\
$+$ & $+$ & $+$ & $-$ & $+$ & $+$ &
$\frac{\varepsilon_{2y}}{\varepsilon_{1y}}<\frac{\mu_{1z}}{\mu_{2z}}$  \\
$+$ & $+$ & $+$ & $+$ & $-$ & $+$ &
$\frac{\mu_{2x}}{\mu_{1x}}>\frac{\varepsilon_{2y}}{\varepsilon_{1y}}$\\
$*$ & $*$ & $*$ & $+$ & $+$ & $-$ & $\times$  \\
$+$ & $+$ & $-$ & $*$ & $*$ & $*$ & $\times$   \\
$+$ & $+$ & $+$ & $+$ & $-$ & $-$ &
$\frac{\varepsilon_{2y}}{\varepsilon_{1y}}>\frac{\mu_{1z}}{\mu_{2z}}$  \\
$+$ & $+$ & $+$ & $-$ & $+$ & $-$ &
$\frac{\mu_{2x}}{\mu_{1x}}<\frac{\varepsilon_{2y}}{\varepsilon_{1y}}$  \\
$+$ & $-$ & $+$ & $+$ & $-$ & $+$ &
$\frac{\mu_{2x}}{\mu_{1x}}<\frac{\varepsilon_{2y}}{\varepsilon_{1y}}\cap
 \frac{\varepsilon_{2y}}{\varepsilon_{1y}}>\frac{\mu_{1z}}{\mu_{2z}}$,
 $\frac{\mu_{2x}}{\mu_{1x}}>\frac{\varepsilon_{2y}}{\varepsilon_{1y}}\cap
 \frac{\varepsilon_{2y}}{\varepsilon_{1y}}<\frac{\mu_{1z}}{\mu_{2z}}$ \\
$+$ & $-$ & $+$ & $-$ & $+$ & $+$ & $\forall$  \\
$-$ & $+$ & $+$ & $-$ & $+$ & $+$ &
$\frac{\mu_{2x}}{\mu_{1x}}<\frac{\varepsilon_{2y}}{\varepsilon_{1y}}\cap
 \frac{\varepsilon_{2y}}{\varepsilon_{1y}}>\frac{\mu_{1z}}{\mu_{2z}}$,
 $\frac{\mu_{2x}}{\mu_{1x}}>\frac{\varepsilon_{2y}}{\varepsilon_{1y}}\cap
 \frac{\varepsilon_{2y}}{\varepsilon_{1y}}<\frac{\mu_{1z}}{\mu_{2z}}$\\
$+$ & $+$ & $+$ & $-$ & $-$ & $-$ &
$\frac{\mu_{2x}}{\mu_{1x}}>\frac{\varepsilon_{2y}}{\varepsilon_{1y}}\cap
 \frac{\varepsilon_{2y}}{\varepsilon_{1y}}>\frac{\mu_{1z}}{\mu_{2z}}$,
 $\frac{\mu_{2x}}{\mu_{1x}}<\frac{\varepsilon_{2y}}{\varepsilon_{1y}}\cap
 \frac{\varepsilon_{2y}}{\varepsilon_{1y}}<\frac{\mu_{1z}}{\mu_{2z}}$  \\
 $+$ & $-$ & $+$ & $+$ & $-$ & $-$ &  $\forall$ \\
$+$ & $-$ & $+$ & $-$ & $+$ & $-$ &
$\frac{\mu_{2x}}{\mu_{1x}}>\frac{\varepsilon_{2y}}{\varepsilon_{1y}}\cap
 \frac{\varepsilon_{2y}}{\varepsilon_{1y}}>\frac{\mu_{1z}}{\mu_{2z}}$,
 $\frac{\mu_{2x}}{\mu_{1x}}<\frac{\varepsilon_{2y}}{\varepsilon_{1y}}\cap
 \frac{\varepsilon_{2y}}{\varepsilon_{1y}}<\frac{\mu_{1z}}{\mu_{2z}}$  \\
 $+$ & $-$ & $-$ & $-$ & $+$ & $+$ &
$\frac{\mu_{2x}}{\mu_{1x}}<\frac{\varepsilon_{2y}}{\varepsilon_{1y}}\cap
 \frac{\varepsilon_{2y}}{\varepsilon_{1y}}>\frac{\mu_{1z}}{\mu_{2z}}$,
 $\frac{\mu_{2x}}{\mu_{1x}}>\frac{\varepsilon_{2y}}{\varepsilon_{1y}}\cap
 \frac{\varepsilon_{2y}}{\varepsilon_{1y}}<\frac{\mu_{1z}}{\mu_{2z}}$  \\
 \end{tabular}
\end{ruledtabular}
{Note: $*$ denotes either of $\pm$, $\forall$ indicates that
Brewster angle exists for any parameter values.}
\end{table}
Note that, if the first medium is vacuum, ${\bf E}^i_{r0}=0$, then
zero-reflection  can occur only when ${\bf E}^t_{r0}=0$. In other
words, the Brewster effect is because the contributions from
electric and magnetic dipoles of the medium to the reflected field
in vacuum add up to zero, which is also pointed out in
Ref.~\cite{Fu2005}. In order to describe these conclusions, we
give numerical examples in Fig.~\ref{figfield}.
We also follow the method in Ref.~\cite{Lu2004} to simulate in
Fig.~\ref{figbeam} a beam ${\bf E}_i={\bf E}_{i0}\int
\hbox{d}k_\perp e^{i({\bf k}_0+{\bf k}_\perp)\cdot {\bf r}}f
({k}_\perp)$ incident from an anisotropic material to another
where $f ({k}_\perp)$ is the Gaussian modulation. We can see
numerical simulations are in agreement with the theoretical
conclusions. Experimentally Brewster effect has been realized with
metamaterials \cite{Tamayama2006}. So,  one can realize zero
reflection through choosing appropriate material parameters, which
may lead to make polarizers or light splitters.

\section{Conclusion}

In summary, we carried out a systematical investigation on the
propagation of wave between two anisotropic magnetoelectric
materials. In terms of Hertz vectors and the principle of
superposition we derive the properties of propagation and
generalize the extinction theorem, so that the propagation between
two arbitrary materials can be investigated in a unified
framework. We apply the results to explain the physical origin of
Brewster effect. The methods do not require boundary conditions,
but can disclose the microscopic process of light propagation in
materials, and avoid complex calculations usually encountered in
using the extinction theorem. So the methods can be applied to
other problems of wave propagation, such as scattering of light
\cite{Kong2001}, propagation through stratified media
\cite{Karam1996} or metamaterials \cite{Belov2006}, and so on.

\begin{acknowledgements}
This work was supported in part by the National Natural Science
Foundation of China (No.~10125521, 10535010) and the 973 National
Major State Basic Research and Development of China (G2000077400).
\end{acknowledgements}



\begin{references}

\bibitem{Born} M. Born and E. Wolf, \textit{Principles of Optics}, 7th ed. (Cambridge, Cambridge, 1999).

\bibitem{Feynman1963} R. P. Feynman, R. B. Leighton, and M. Sands, \textit{The Feynman Lectures
on Physics} (Addison-Wesley, 1963), Vol. 1, Secs. 31 and 30-7.

\bibitem{Wolf1972}  E. Lalor and E. Wolf, {J. Opt. Soc. Am.} {\bf 62}, 1165 (1972).

\bibitem{Sein1972}  J. L. Birman and J. J. Sein, \prb {\bf 6}, 2482 (1972).

\bibitem{Fearn1996} H. Fearn, D. F. V. James, and P. W. Milonni,  Am. J. Phys. {\bf 64}, 986 (1996).

\bibitem{Reali1982}  G. C. Reali, {J. Opt. Soc. Am.} {\bf 72}, 1421 (1982).

\bibitem{Lai2002} H. M. Lai, Y. P. Lau, and W. H. Wong,  Am. J. Phys. {\bf 70}, 173¨C179 (2002).

\bibitem{Fu2005}  C. Fu, Z. M. Zhang, and P. N. First, Applied Optics. {\bf 44}, 3716 (2005).

\bibitem{Kong2001}  L. Tsang, J. A. Kong, and K. H. Ding, \textit{Scattering of Electromagnetic Waves: Theories and
Applications} (New York, John Wiley \& Sons Inc, 2000).

\bibitem{Veselago1968}  V. G.  Veselago, Sov. Phys. Usp. {\bf 10},  509 (1968).

\bibitem{Shelby2001} R. A. Shelby, D. R. Smith, and S. Schultz, Science {\bf 292},
77 (2001).

\bibitem{Pendry2000} J. B. Pendry,  \prl {\bf 85}, 3966 (2000).

\bibitem{Lindell2001} I. V. Lindell, S. A. Tretyakov, K. I. Nikoskinen, and S. Ilvonen,
                      Microw. Opt. Technol. Lett. {\bf 31}, 129 (2001).

\bibitem{Smith2004}  D. R. Smith, P. Kolinko, and D. Schurig, {J. Opt. Soc. Am. B} {\bf 21},
1032 (2004).

\bibitem{Luo2002}  C. Luo, S. G. Johnson, J. D. Joannopoulos, and J. B. Pendry, {Optics Express} {\bf 11}, 746 (2003).

\bibitem{Hu2002}  L. B. Hu, S. T. Chui, \prb {\bf 66}, 085108 (2002).

\bibitem{Lu2004}  W. T. Lu, J. B. Sokoloff, and S. Sridhar, \pre {\bf 69}, 026604 (2004).

\bibitem{Lakhtakia2004}  Tom G. Mackay and A. Lakhtakia, \pre {\bf 69}, 026602 (2004)

\bibitem{Lakhtakia2006}  R. A. Depine, M. E. Inchaussandague, A. Lakhtakia, {J. Opt. Soc. Am. A.} {\bf 23}, 949 (2006).

\bibitem{Luo2005} H. Luo, W. Hu, X. Yi, H. Liu, and J. Zhu, {\oc} {\bf 254},
353  (2005).

\bibitem{Luo2006a}  H. Luo, W. Hu, W. Shu, F. Li and Z. Ren, Europhys. Lett. {\bf 74}, 1081 (2006).

\bibitem{Luo2006b}  H. Luo, W. Shu, F. Li, and Z. Ren, {\oc}, in press (2006).

\bibitem{Shen2006}  N. H. Shen, Q. Wang,  J. Chen, Y. X. Fan, J. P. Ding, H. T. Wang,
 {J. Opt. Soc. Am. B.} {\bf 23}, 904 (2006).

\bibitem{Zhou2003}  L. Zhou, C. T. Chan,  and P. Sheng, \prb {\bf 68}, 115424 (2003).

\bibitem{Grzegorczyk2005} T. M. Grzegorczyk, Z. M. Thomas, and J. A. Kong,  \apl {\bf 86}, 251909 (2005).

\bibitem{Tamayama2006}  Y. Tamayama, T. Nakanishi, K. Sugiyama, and M. Kitano, \prb {\bf 73}, 193104  (2006)

\bibitem{Kong2000} J. A. Kong, \textit{Electromagnetic wave theory} (EMW, New York, 2000).

\bibitem{Karam1996}  M. A. Karam, {J. Opt. Soc. Am. A} {\bf 13}, 2208 (1996).

\bibitem{Belov2006}  P. A. Belov, \prb {\bf 73}, 045102 (2006).
\end{references}
\end{document}